\documentclass[reqno]{amsart}

\makeatletter
\@addtoreset{equation}{section}

\makeatother

\newcommand{\ba}{\[\begin{aligned}}
\newcommand{\ea}{\end{aligned}\]}

\usepackage{amsmath}
\usepackage[foot]{amsaddr}
\usepackage{amsthm,amssymb,mathrsfs}
\usepackage{graphicx}
\usepackage{natbib}
\usepackage{color}

\theoremstyle{remark}

\title[]{Nondestructive assessment of ripeness in kiwifruit with near-infrared pulse illumination}

\author[]{Hiyori Ishiji, Hiroki Kanatsu, Masaki Komatsubara, Shingo Minata, Masaki Uesugi, Kohei Yuguchi, and Manabu Machida}
\address{Department of Informatics, Faculty of Engineering, Kindai University, Higashi-Hiroshima 739-2116, Japan}
\email{machida@hiro.kindai.ac.jp}

\begin{document}

\begin{abstract}
We investigate the ripeness of kiwifruit nondestructively with near-infrared pulse illumination. With this measurements in time domain, only one frequency (the wavelength $800\,{\rm nm}$ is required. Measurements were performed on three golden kiwifruits over a period of ten days. To quantify changes in temporal profiles of the detected light, we introduce two indices: the relative ripeness $r(n)$ and the first Wasserstein distance $W_1(n)$. Both indices exhibit nonmonotonic behavior.
\end{abstract}

\maketitle

\section{Introduction}
\label{intro}

A conventional way of using near-infrared light for nondestructive assessment of foods is to illuminate the foods with steady light. Here we consider a pulse illumination against kiwifruit.

The history of the nondestructive assessment for foods with near-infrared light goes back to 1950's. After initial research \citep{Norris55,Norris58}, this methodology was brought to practical application for meat \citep{Ben-Gera-Norris68} and for fruits \citep{Kawano-Watanabe-Iwamoto92,Kawano-Fujiwara-Iwamoto93}.

Near-infrared light penetrates biological tissue undergoing multiple scattering. Light propagation in biological tissue is characterized by absorption and scattering. It was shown that light propagation in kiwifruit obeys the radiative transport equation \citep{Baranyai-Zude09} and the diffusion equation \citep{Valero-etal04} depending on the propagation distance. When light is sent from the surface, the penetration depth depends on the wavelength of light. See \citep{Ma-etal21} for the dependence of light reflectance on wavelengths between $1000\,{\rm nm}$ and $2300\,{\rm nm}$. Also, see a recent review \citep{Anjali-etal24} for conventional measurements with multiple wavelengths.

It is desirable to use the most suitable wavelength to assess ripeness in food products. In the case of kiwifruit, suitable wavelengths lie between $750\,{\rm nm}$ and $900\,{\rm nm}$ \citep{Valero-etal04}. However, sufficient information cannot be obtained with a constant illumination of a single-frequency light. Ma et al., detected light not as a function of frequencies but as a function of time by sending a pulse of near-infrared light \citep{Ma-etal22}. See \citep{Tsuchikawa-etal02} for the use of time-of-flight near-infrared spectroscopy. 

In this paper, we consider nondestructive assessment of ripeness in kiwifruit with near-infrared pulse illumination of a single wavelength.

\section{Material and methods}
\label{matemetho}

\subsection{Measurements}

\begin{figure}[ht]
\centering
\includegraphics[width=0.3\textwidth]{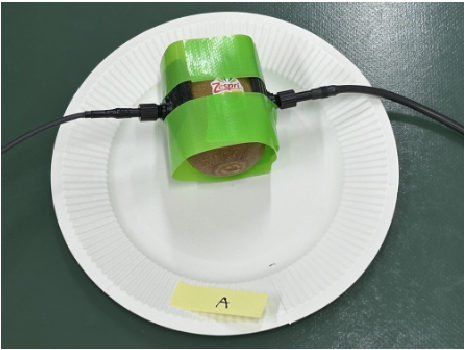}
\caption{
Kiwi A.
}
\label{fig:kiwiA}
\end{figure}

\begin{figure}[ht]
\centering
\includegraphics[width=0.3\textwidth,angle=270]{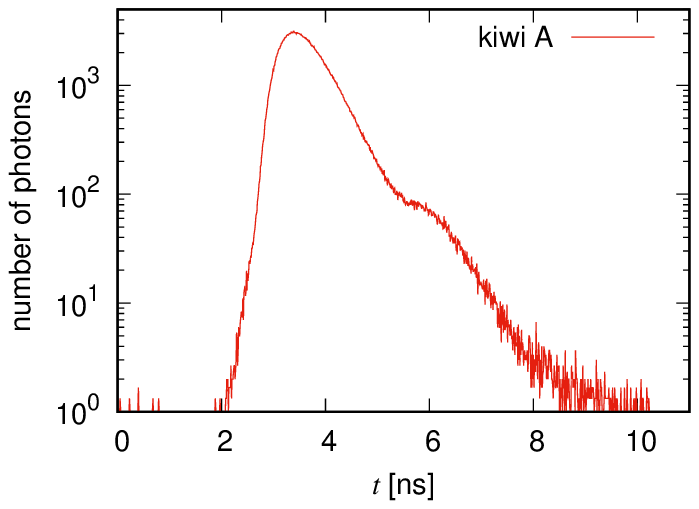}
\caption{
Detected near-infrared light for Kiwi A.
}
\label{fig:kiwiA_time}
\end{figure}

Three golden kiwifruits (kiwi A, kiwi B, and kiwi C) were prepared. Two optical fibers (one is for illumination and the other is for detection) were attached to the kiwifruits as shown in Fig.~\ref{fig:kiwiA}. The two optical fibers were connected to TRS (Time-Resolved Spectroscopy, Hamamatsu Photonics), which sends and detects near-infrared light (wavelength $800\,{\rm nm}$). An example of detected light (Kiwi A) is shown in Fig.~\ref{fig:kiwiA_time}.

\subsection{Preprocessing}

Let $I_j(t_k;n)$ be the intensity of light (number of photons) detected at time $t$ on the $2n$th day ($n=0,1,\dots,N$; we will set $N=5$). Each time, we performed three consecutive measurements ($j=1,2,3$). The intensity $I_j(t_k;n)$ is measured every $\Delta t$. That is,
\begin{equation}
t_k=k\Delta t,\quad k=0,1,\dots,M.
\end{equation}
We have $\Delta t=10\,{\rm ps}$ and $M=1023$. This means that the observation time is $10\,{\rm ns}$.

After experiments, we took the average of three $I_j(t;n)$:
\begin{equation}
I(t_k;n)=\frac{1}{3}\left(I_1(t_k;n)+I_2(t_k;n)+I_3(t_k;n)\right).
\end{equation}
In Fig.~\ref{fig:kiwiA_time}, $I(t_k;0)$ for Kiwi A is shown. Then we took the moving average using $2m+1$ points (we set $m=2$):
\begin{equation}
\bar{I}(t_k;n)=\frac{1}{2m+1}\sum_{l=-m}^mI(t_{k+l};n),\quad
m\le k\le M-m.
\end{equation}
Next, we normalize $\bar{I}(t_k;n)$ as
\begin{equation}
p_n(t_k)=\frac{\bar{I}(t_k;n)}{\Delta t\sum_{k=m}^{M-m}\bar{I}(t_k;n)},\quad 
m\le k\le M-m.
\end{equation}
We note that $\Delta t\sum_{k=m}^{M-m}p_n(t_k)=1$.

\subsection{Squared intensity displacement}

To explore how the intensity changes over time, we subtract $p_0(t_k)$ from other intensities:
\begin{equation}
d_n(t_k)=p_n(t_k)-p_0(t_k),\quad n=1,\dots,N,\quad m\le k\le M-m.
\end{equation}
We call $d_n(t_k)^2=d_n^2$ squared intensity displacement.

Furthermore, we introduce $r(n)$ below to quantify ripeness relatively from the $0$th day:
\begin{equation}
r(n)=\Delta t\sum_{k=m}^{M-m}d_n(t_k)^2.
\end{equation}
We call $r(n)$ the relative ripeness. Note that $r(n)$ is the area of $d_n(t_k)^2$.

\subsection{Wasserstein distance}

Since $\Delta t$ is small and $m$ is negligibly small compared with $M$, we write $p_n(t_k)$ ($m\le k\le M-m$) as $p_n(t)$ ($0<t<T$), where $T=M\Delta t$. Seeking for an alternative index for ripeness, let us consider the cumulative distribution function, which is given by
\begin{equation}
F_n(x)=\int_0^xp_n(t)\,dt,\quad 0<x<T
\end{equation}
for $n=0,\dots,N$. We note that $F_n$ is continuous (as long as $t_k$ can be approximately replaced by $t$) and strictly monotonically increasing. The quantile (the inverse of the cumulative density function) is given by
\begin{equation}
F_n^{-1}(s)=\inf\left\{x\in(0,T);\;F(x)\ge s\right\}.
\end{equation}
We introduce the $p$th Wasserstein distance as \citep{Villani03}
\begin{equation}
W_p(n)=\left(\int_0^1\left|F_n^{-1}(s)-F_0^{-1}(s)\right|^p\,ds\right)^{1/p}
\end{equation}
for $n=1,\dots,N$. In particular,
\begin{align}
W_1(n)
&=
\int_0^1\left|F_n^{-1}(s)-F_0^{-1}(s)\right|\,ds
\nonumber \\
&=
\int_0^T|F_n(x)-F_0(x)|\,dx.
\end{align}

\section{Results}
\label{results}

\begin{figure*}[ht]
\centering
\includegraphics[width=0.3\textwidth,angle=270]{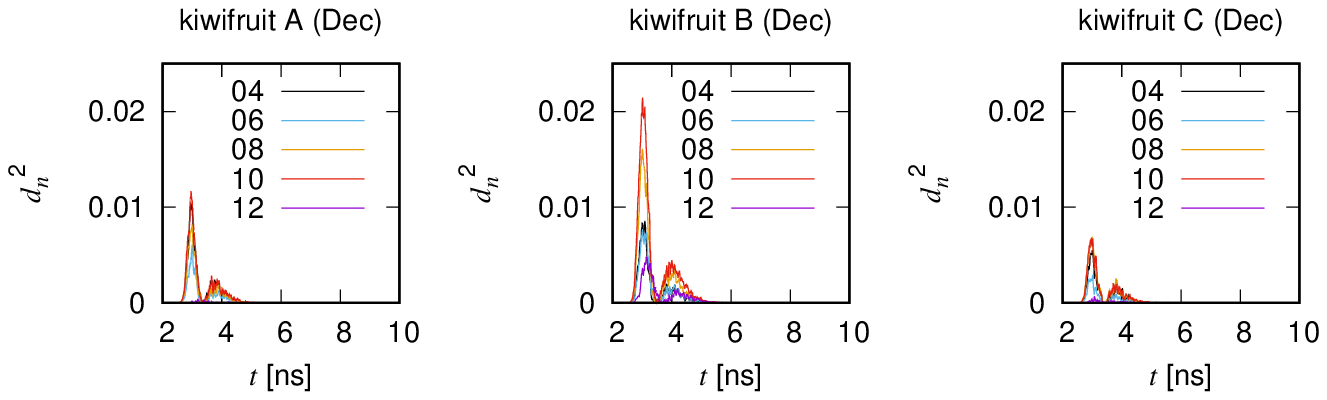}
\caption{
For kiwis A, B, C, $d_n(t_k)^2$ are plotted for Dec 4 ($n=1$), Dec 6 ($n=2$), Dec 8 ($n=3$), Dec 10 ($n=4$), and Dec 12 ($n=5$).
}
\label{fig:method3_Dec}
\end{figure*}

\begin{figure}[ht]
\centering
\includegraphics[width=0.3\textwidth,angle=270]{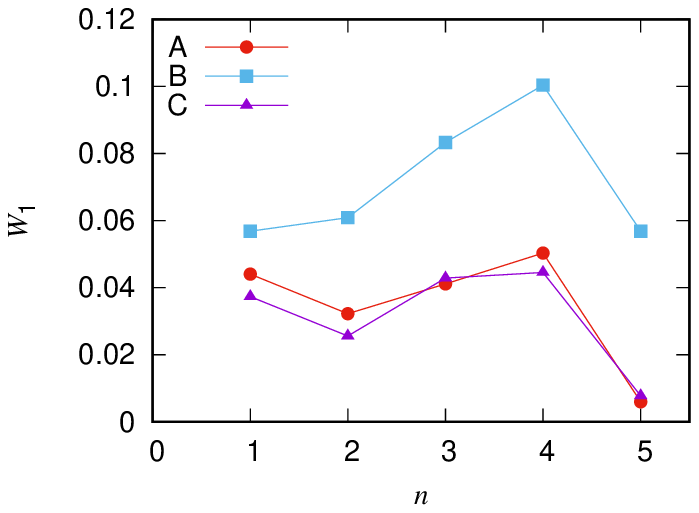}
\caption{
The $1$st Wasserstein distance $W_1$ for $n=1$ (Dec 4), $n=2$ (Dec 6), $n=3$ (Dec 8), $n=4$ (Dec 10), and $n=5$ (Dec 12).
}
\label{fig:wass1}
\end{figure}

Figure \ref{fig:method3_Dec} shows $d_n(t_k)^2$ from experiments in December 4th through 12th in 2025 for three samples of kiwifruit, to which holder were directly attached. For each sample of the kiwifruits, $r(n)$ is summarized in Table \ref{table:r_kiwi}. In Fig.~\ref{fig:wass1}, $W_1(n)$ is plotted.

\begin{table*}[ht]
\centering
\begin{tabular}{l|ccccc}
& $n=1$ (Dec 4) & $n=2$ (Dec 6) & $n=3$ (Dec 8) & $n=4$ (Dec 10) & $n=5$ (Dec 12)\\ \hline
kiwifruit A & 0.00404 & 0.00208 & 0.00328 & 0.00461 & 0.00009 \\
kiwifruit B & 0.00392 & 0.00388 & 0.00777 & 0.01069 & 0.00269 \\
kiwifruit C & 0.00241 & 0.00120 & 0.00305 & 0.00316 & 0.00017
\end{tabular}
\caption{
relative ripeness
}
\label{table:r_kiwi}
\end{table*}

\section{Discussion}
\label{discussion}

We found that each $d_n$ ($n=1,\dots,5$) has two peaks in Fig.~\ref{fig:method3_Dec}. Since light in kiwifruit is known to obey the diffusion equation \citep{Valero-etal04}, from general theory of near-infrared spectroscopy \citep{Hielscher-etal95}, the left peak is related to the scattering of light in the food and the right peak reflects the absorption of light in the food. We can infer that both scattering and absorption decrease inside the kiwifruit as time passes.

The behavior of $d_n$ ($n=1,\dots,5$) was not monotonic. This tendency is quantitatively seen in $r(n)$ in Table \ref{table:r_kiwi}. Moreover, the nonmonotonic time-evolution is also seen in $W_1(n)$ in Fig.~\ref{fig:wass1}.

\section{Conclusion}
\label{concl}

Ripeness of kiwifruit was investigated with time-resolved near-infrared measurements. We found that the relative ripeness $r(n)$ and $W_1$ change nonmonotonically in time.

\section*{Declaration of Competing Interest}
The authors declare that they have no known competing financial interests or personal relationships that could have appeared to influence the work reported in this paper.

\section*{Acknowledgments}
This research was partially supported by Satake Science and Technology Foundation (Higashi-Hiroshima, Japan). The experiments were performed in the Fundamental Technology for Next Generation Research Institute at Kindai University.

\section*{Data availability}
Data will be made available on request.


\end{document}